\documentclass[prl,twocolumn,amsmath,amssymb,pre,array]{revtex4}
\usepackage[dvips]{graphicx}
\usepackage{epsfig}
\usepackage{psfrag}

\begin{document}
\title{Magnetic Quantum Oscillations of the Conductivity in
Two-dimensional Conductors with Localization}
\author{V. M. Gvozdikov}
\affiliation{Max-Planck-Institut f\"ur Physik komplexer Systeme, N\"othnitzer
Strasse 38, D-01187 Dresden, Germany}
\affiliation{Kharkov National University, 61077, Kharkov, Ukraine}
\date{\today}

\begin{abstract}
An analytic theory is developed for the diagonal conductivity
$\sigma_{xx}$ of a 2D conductor which takes account
of the localized states in the broaden Landau levels.
In the low-field region $\sigma_{xx}$ display the Shubnikov-de Haas
oscillations which in the limit $\Omega \tau\gg 1$ transforms into
the sharp peaks ($\Omega$ is the cyclotron frequency,
$\tau$ is the electron scattering time). Between the peaks
$\sigma_{xx}\to 0$.
With the decrease of
temperature, $T$, the peaks in $\sigma_{xx}$ display first a thermal
activation behavior $\sigma_{xx}\propto \exp(-\Delta/T)$, which then
crosses over into the variable-range-hopping regime
at lower temperatures with $\sigma_{xx}\propto 1/T \exp(-\sqrt{T_{0}/T})$
(the prefactor $1/T$ is absent in the conductance).

PACS numbers: 73.43.-f, 73.40.Gk, 75.47.-m
\end{abstract}

\maketitle

Despite more than two decades of intensive studies, some open questions
remain in the quantum magnetic oscillations of the 2D conductors.
Even for the most studied case of the integer quantum Hall
effect (IQHE) a coherent description is absent for different field
and temperature regimes observed in the diagonal conductivity $\sigma_{xx}$ 
\cite{1}.
In particular, it is not clear so far why quantum oscillations in
$\sigma_{xx}$ survive in spite of that most states within the broaden
Landau levels (LL) are localized (i); Why
$\sigma_{xx}\to 0$ between the peaks in the limit $\Omega \tau\gg 1$,
if at low fields it displays a standard Shubnikov-de Haas (SdH) 
oscillations (ii); Why with the decrease of
temperature, $T$, the peaks in $\sigma_{xx}$ display first a thermal
activation behavior $\sigma_{xx}\propto \exp(-\Delta/T)$, which then
crosses over into the variable-range-hopping (VRH) regime
at low temperatures with $\sigma_{xx}\propto 1/T \exp(-\sqrt{T_{0}/T})$ 
(iii); Why the prefactor $1/T$ is absent in the conductance (iv).

The localization in the IQHE picture plays a crucial role. It is believed
that extended states are at the center of the broaden LL and
all the other states are localized \cite {1}. At high fields the localized
states mean that Landau orbits drift along the closed equipotential
contours of the impurity potential. At places
where contours come close electrons can tunnel from one contour
to another providing thereby a conductivity mechanism through
the extended states. The diagonal conductivity $\sigma_{xx}$ and the
Hall conductivity $\sigma_{xy}$ are closely related in the IQHE. The
peaks in the $\sigma_{xx}$ are exactly at the same fields where
$\sigma_{xy}$ transits from one plateau to another. The ideal picture
of the IQHE at $T=0$ assumes that $\sigma_{xx}=0$ within the plateaus
while the $\sigma_{xy}=ne^2/h$ is quantized ($n$ is an integer).
In real experiments at $T\ne 0$ the $\sigma_{xx}\ne 0$ within
the plateaus in the low-field region and displays the SdH
oscillations at smaller fields in which plateaus in the
$\sigma_{xy}$ are unresolved.
 
The purpose of this paper is to study analytically the quantum magnetic
oscillations in the $\sigma_{xx}$ in the 2D
conductor with the localized states in the broaden LLs and
to proof the properties (i)-(iv).

The conductivity due to the electron tunneling between the Landau orbits
was calculated in \cite{2} for the case of incoherent hopping across the
layers of a layered conductor. This hopping mechanism remains
in effect if Landau orbits lay within the same conducting plane or belong
to the tunnel-coupled 2D conductors. The latter was proved by a
recent observation of the typical IQHE behavior in the tunneling
conductance of a two coupled Hall bars \cite{3}.
According to \cite{2} the tunneling SdH
conductivity can be written as a sum of the Boltzmann
($\sigma _{B}$) and quantum ($\sigma _{Q}$) terms:
$\sigma _{xx}=\sigma _{B}+\sigma _{Q}$ , where
\begin{equation}
\sigma _{B}=\sigma _{0}\int d\varepsilon \frac{dE}{
\pi }g(\varepsilon )v_{x}^{2}(\varepsilon)
\left(-\frac{\partial f}{\partial E}\right)
\tau S[\lambda ,\delta (E,\varepsilon )],
\label{1}
\end{equation}
\begin{equation}
\sigma _{Q}=\sigma _{0}\int d\varepsilon \frac{dE}{
\pi }g(\varepsilon )v_{x}^{2}(\varepsilon)
\left(\frac{\partial f}{\partial E}\right)
\frac{2\pi }{\Omega }\frac{\partial }{\partial \lambda }
S[\lambda ,\delta (E,\varepsilon )].
\label{2}
\end{equation}
Here $\lambda (E) =2\pi/\Omega \tau,\delta (E,\varepsilon) =2\pi
(E+\varepsilon)/\hbar \Omega$,
$\sigma _{0}=e^2 N_{L}/\Omega$, $N_{L}=\Phi/S\Phi _{0}$ is
the electron density at the LL, $\Phi$ is the flux
through a sample, $\Phi _{0}=\hbar c/2\pi e$, and
\begin{equation}
S(\lambda ,\delta )=\sum_{p=-\infty }^{\infty }(-1)^{p}e^{-\left| p\right| \lambda }\cos p\delta
=\frac{\sinh \lambda }{\cosh \lambda +\cos \delta }.
\label{3}
\end{equation}
The variable $\varepsilon$ describes the LLs
broadening by impurities with the density of states (DOS)
$g(\varepsilon)$:
\begin{equation}
E_{n}(\varepsilon)=\hbar \Omega (n+1/2)+\varepsilon.
\label{4}
\end{equation}
The electron
velocity $v_{x}$ is related to the tunneling matrix elements by\cite{2}
\begin{equation}
v_{x}(\varepsilon)= \frac{|t_{\varepsilon,\varepsilon}|R}{\hbar\sqrt{2}}
\label{5}
\end{equation}
where $R$ and $\hbar\sqrt{2}/|t_{\varepsilon,\varepsilon}|$ are 
correspondingly the distance and the time of the tunneling.
The strong point of the above equations is that we can learn much about
the $\sigma _{xx}(B,T)$ without resort to the specific models for the
localization ($B$ is the magnetic field). In any such model the $g(\varepsilon )$ has
a narrow band of delocalized states where the
$v_{x}(\varepsilon)\ne 0$. It is generally accepted now that only one
state, precisely at the LL ($\varepsilon=0$) is
delocalized. For the localized states $v_{x}(\varepsilon)= 0$.
Thus, only one level $\varepsilon=0$, or a small stripe
of delocalized states, contribute into Eqs.(\ref{1}),(\ref{2}).

The scattering time $\tau$ in general is a model-dependent function of
the energy which is inversely proportional to the scattering
probability for the conducting (delocalized) electrons. The latter
belong to a narrow stripe in the $g(\varepsilon )$
while the rest of electrons are localized and produce a reservoir
of states stabilizing oscillations in $\tau$. Besides, only
$\varepsilon =0$ contribute into the $\sigma _{xx}$. Thus, we can
put $\tau$=const. in Eqs.(\ref{1}),(\ref{2}) which yields:
\begin{equation}
\sigma _{xx}=\sigma _{\tau }
\int \frac{dE}{\pi }\left(-\frac{\partial f}{\partial E}\right)
[G_{B}(\lambda ,E)+G_{Q}(\lambda ,E)],
\label{6}
\end{equation}
\begin{equation}
G_{B}(\lambda,E)=
S[\lambda ,\Delta(E)],
\label{7}
\end{equation}
\begin{equation}
G_{Q}(\lambda,E)=-\lambda \frac{\partial }{\partial \lambda }
S[\lambda ,\Delta(E)]=
-\lambda \frac{1+\cosh \lambda \cos \Delta }
{\left( \cosh \lambda +\cos\Delta \right) ^{2}},
\label{8}
\end{equation}
where $\Delta(E)=2\pi E/\hbar \Omega$ and
\begin{equation}
\sigma _{\tau}=\frac{e^2 N_{L}\tau <v_{x}^{2}>}{\hbar\Omega}.
\label{9}
\end{equation}
The average of the velocity squared, is given by
\begin{equation}
<v_{x}^{2}> =\frac{R^2}{2\hbar^2}\int^{\varepsilon_{max}}_{\varepsilon_{min}}
d\varepsilon g(\varepsilon)|t_{\varepsilon,\varepsilon}|^{2}.
\label{10}
\end{equation}
Integral in Eq.(\ref{10}) is taken within the narrow stripe of the
delocalized states.
\begin{figure}[htb]
\centering
{\bf(a)}
\includegraphics[width=0.35\textwidth]{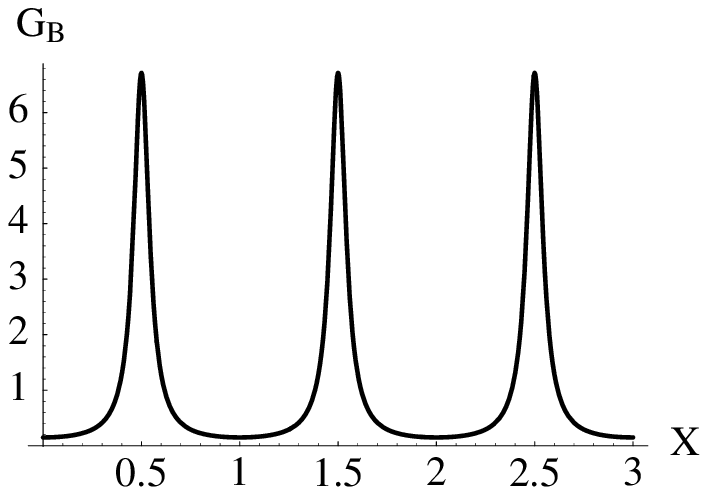}\\
\centering
{\bf(b)}
\includegraphics[width=0.35\textwidth]{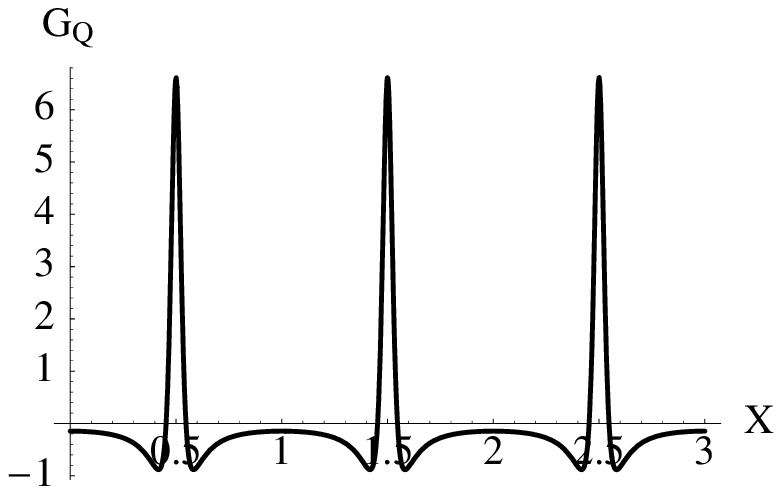}\\
\centering
{\bf(c)}
\includegraphics[width=0.35\textwidth]{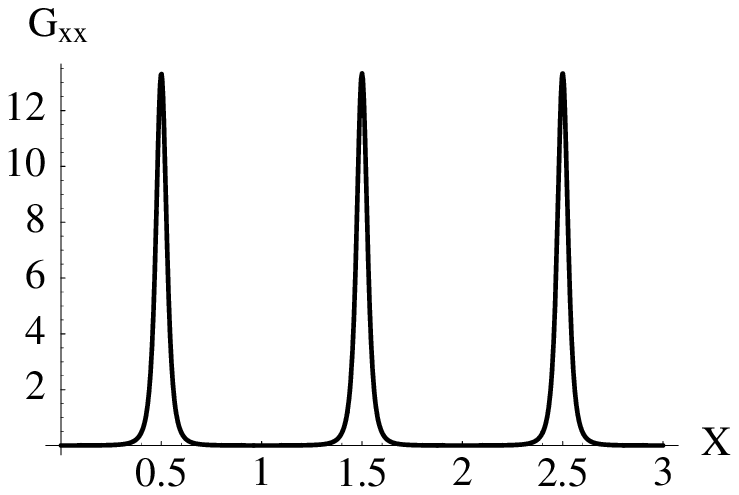}\\
\caption{The Boltzmann, $G_{B}= S(\lambda, 2\pi X )$ (Fig.1a), and
the quantum, $G_{Q}=-\lambda \frac{\partial }{\partial \lambda }S(\lambda ,2\pi X )$
(Fig.1b), contributions into the
conductivity $\sigma_{xx}$ in Eq.(\ref{6}), and their sum
$G_{xx}=G_{B}+G_{Q}$ (Fig.1c). $ X = E /\hbar\Omega $, $\lambda = 0.3 $.}
\end{figure}
The functions $G_{B}(\lambda ,E)$ and $G_{Q}(\lambda ,E)$ are sharply
peaked at the LLs $E=E_{n}$ and between them they nearly
compensate each other, as one can see in Fig.1. This important point
demonstrates clearly that the Boltzmann term alone, $G_{B}(\lambda ,E)$,
is insufficient for the correct description and
only by taking account of the quantum term, $G_{Q}(\lambda ,E)$, one can
explain why $\sigma _{xx}$ tends to zero between the peaks in the IQHE.
The width of the peaks in Fig.1 in the energy scale is of the order of
$\hbar/\tau$. If the temperature $ T>>\hbar/\tau $, then the peaked
function $-(\partial f/\partial E)$ is broader than, the
$G_{xx}(\lambda,E)=G_{B}(\lambda,E)+G_{Q}(\lambda,E)$, and we can
approximate the $G_{xx}(\lambda,E)$ in Eq.(\ref{6}) by
\begin{equation}
G_{xx}(\lambda,E)\approx \frac{2}{\pi }
\sum_{n=-\infty }^{\infty }\frac{\eta }{(n+1/2-E/\hbar \Omega)^{2}+
\eta ^{2}}
\label{11}
\end{equation}
where $\eta=\lambda/2\pi$. For $\eta\ll 1$ Eq.(\ref{11}) can
be easily proved analytically with the help of the identity \cite{2}
\begin{equation}
\frac{1}{\pi }\sum_{p=-\infty }^{\infty }\frac{\eta }{(n+a)^{2}+\eta ^{2}}
=\frac{\sinh 2\pi \eta}{\cosh 2\pi \eta -\cos 2\pi a}.
\label{12}
\end{equation}
Thus, for high temperatures, $T>>\hbar/\tau$, we have
\begin{equation}
\sigma_{xx}(B)\approx\sigma_{\tau}\frac{\hbar \Omega }{4\pi T}\sum_{n}
\cosh ^{-2}\left( \frac{E_{n}-\mu }{T}\right).
\label{13}
\end{equation}
This sharply-peaked function of the $\hbar\Omega$ is shown in Fig.2.
The same function describes the quantum magnetic oscillations of
the ultrasound absorption in metals \cite{4}.
\begin{figure}[htb]
\centering
\includegraphics[width=0.35\textwidth]{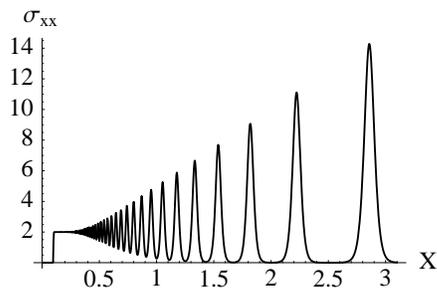}\\
\centering
\caption{The conductivity $\sigma_{xx}$ [see Eq.(\ref{13})] in units of
$\sigma_{\tau}$ as a function of the $X=\hbar\Omega$. The conventional 
energy units accepted in which $T=0.2$ and $E_{F}=10$.}
\end{figure}
A temperature dependence of the peaks in $\sigma_{xx}(B)$ for different
temperatures is plotted in Fig.3.
\begin{figure}[htb]
\centering
\includegraphics[width=0.35\textwidth]{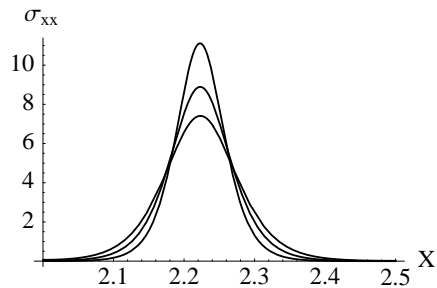}\\
\centering
\caption{The same as in Fig.2 for three different temperatures
$T=0.2$, $T=0.25$, and $T=0.3$ (from top to bottom).}
\end{figure}
Under the condition $\hbar \Omega /T>>1$, the conductivity
$\sigma_{xx}$ at the maxima (i.e. when $E_{n}=\mu $) is given by
$\sigma_{xx}=\sigma _{\tau}\frac{\hbar \Omega }{4\pi T}$.
At the minima (i.e. when the chemical potential $\mu $ falls between the
LL) the conductivity $\sigma_{xx}$ is
exponentially small: $\sigma_{xx}=\sigma _{\tau }
\frac{\hbar \Omega }{4\pi T}\exp \left
( -\frac{\hbar \Omega -E_{0}}{T}\right)$
($E_{0}$ is a position of the $\mu$ between the LL). Such an activation
dependence is well established in the $\sigma_{xx}(T)$ in the IQHE regime 
\cite{1}. At lower
temperatures, $T<<\hbar /\tau$, one can approximate the
$(-\partial f/\partial E)$ by $\delta(E-\mu),$ to obtain
\begin{equation}
\sigma_{xx}\approx \sigma_{\tau }\left(G_{B}[\lambda ,\Delta(\mu)] +
[G_{Q}\lambda ,\Delta(\mu)]\right).
\label{14}
\end{equation}
The $\sigma_{xx}$ in (\ref{14}) is sharply peaked function 
of the $\Delta(\mu)=2\pi\mu/\hbar\Omega$ as shown in Fig.1. The Boltzmann 
and the quantum terms in Eq.(\ref{14}) nearly compensate each other
between the peaks which in the limit $\eta\to 0$ become
narrow Lorentzians of Eq.(\ref{11}). The temperature dependence of the
$\sigma_{xx}$ at $T<<\hbar /\tau$ in Eq.(\ref{14})
comes only from the $\sigma_{\tau}$ due to the VRH mechanism.
The VRH concept in the IQHE problem is now well established \cite{5,6}.
It was introduced
in \cite{7} and well describes the scaling properties of the peaks in the $\sigma_{xx}$ within the
plateau-to-plateau transition region. The diagonal and the Hall 
conductivities in this region are related by the
"semicircle" law \cite{8} . In samples with the mobility of the
order $10^2$Vs at a few tens mK the best experimental fit yields 
\cite{5,6}:
\begin{equation}
\sigma_{\tau}=\frac{A}{T} \exp(-\sqrt{T_{0}/T}), \qquad
T_{0}=C \frac{e^2}{\epsilon\xi}.
\label{15}
\end{equation}
The characteristic temperature $T_{0}$ is proportional to the
Coulomb energy at the localization length $\xi(\nu)$ and $\epsilon$ is
the dielectric constant, $C\sim 1$.
Many experiments and numerical calculations testify in favor of
a univertsal critical behavior of the localization length 
$\xi(\nu)\propto|\nu-\nu_{c}|^{-\gamma}$ near the Landau levels 
\cite{1,5,6,7}.
Here $\nu=N\Phi_{0}/B$ is the filling factor, $\nu_{c}$ is the
critical filling factor, and $\gamma \approx 2.35$ is a universal
critical exponent. The divergency of the $\xi(\nu)$
at $\nu_{c}$ means that this is a critical point for the transition 
from the dielectric to the conducting state.

Eq.(\ref{15}) directly follows from Eqs.(\ref{9}), and (\ref{10}).
In the spirit of the VRH approach, we can estimate the 
$|t_{\varepsilon,\varepsilon}|^{2}$ as proportional to the the
electron hopping probability between the two 1D closed equipotential
impurity-potential-contours at which Landau orbitals are localized.
If $R$ is a distance of the hopping, then
\begin{equation}
|t_{\varepsilon,\varepsilon}|^{2} \propto \exp \left[-\left(
\frac{1}{RN(0)T}+ \frac{2R}{\xi}\right)\right].
\label{16}
\end{equation}
Here we take account of the thermal activation which
helps the tunneling if the initial and final levels are within the
energy stripe of the order of $1/RN(0)$, where $N(0)$ is the DOS at the
Fermi level. Thus, the optimal hopping distance is $R=\sqrt{\xi/2N(0)T}$.
Putting this value into Eqs.(\ref{10}),(\ref{16}) we have
$<v_{x}^{2}>\propto 1/T \exp(-\sqrt{T_{0}/T})$ which, in view of
Eq.(\ref{9}) result in the VRH conductivity given by Eq.(\ref{15}).
The VRH concept was originally applied to the problem of the 
conductivity peak broadening $\Delta\nu$ in \cite{7}. It was shown that 
the temperature,
current, and frequency dependencies of the $\Delta\nu$ can be well
described within this paradigm. Here we derived a prefactor $A/T$ which
also have been observed in the conductivity $\sigma_{xx}(T)$ \cite{5,6}.
However, it should be noted that the prefactor $A/T$ is absent
in the experiments in which a conductance was measured \cite{9,10}.
The difference is because the conductivity in Eq.(\ref{1}),(\ref{2}) is
proportional to the $v_{x}^2\propto R^2\propto 1/T$. The conductance
$\sigma^c_{xx}(T)\propto (e^2/\hbar) |t_{\varepsilon,\varepsilon}|^2$ 
and does not contain a factor $R^2\propto 1/T$. Therefore,
at the same conditions as in Eq.(\ref{14}) the conductance is:
\begin{equation}
\sigma^c_{xx}\approx \sigma^c_{\tau }\left(G_{B}[\lambda ,\Delta(\mu)] +
G_{Q}[\lambda ,\Delta(\mu)]\right),
\label{17}
\end{equation}
\begin{equation}
\sigma^c_{\tau}= A_{c}\exp(-\sqrt{T_{0}/T}).
\label{18}
\end{equation}
Since $T_{0}\propto 1/\xi\propto |\nu -\nu_{c}|^{\gamma}$ the function
$\sigma^c_{\tau}(\nu)$ has a fixed maximum value $\sigma^c_{\tau}= A_{c}$
at $\nu =\nu_{c}$ for different temperatures. This remarkable property
of the conductance is firmly established in
the VRH regime at low temperatures \cite{8,9,10}.
So far we assumed that the chemical potential is a constant.
In 2D conductors $\mu(B)$ is an oscillating function \cite{11} 
satisfying the equation \cite{2}:
\begin{equation}
\mu=E_{f}\pm \frac{\hbar\Omega}{\pi}\arctan\left[\frac{\sin (2\pi \mu/\hbar\Omega)}
{e^{\nu}+\cos (2\pi \mu/\hbar\Omega)} \right].
\label{19}
\end{equation}
The sign (-) here stands for the direct and (+) for the inverse
sawtooth. The amplitude of these oscillations is of the order of the
$\hbar\Omega$ which is small compared to the $E_{F}$.
It was shown in \cite{2} that in a quasi 2D layered
conductor the peaks in the magnetic conductivity across the layers
are split in the case $\mu(B)$ is an inverse sawtooth function. The very
same effect holds for the $\sigma_{xx}$, as shown in Fig.4, which 
displays the
$\sigma_{xx}(B)$ according to Eq.(\ref{14}) with the $\mu(B)$ given by
Eq.(\ref{19}). We also take account of the spin-splitting which is easy
to incorporate by the substitution $\mu\to \mu\pm \mu_{e}B$ into the 
right-hand-side of Eq.(\ref{19}) and by average it over two spin
configurations ($\mu_e$ is the magnetic moment of electron).
\begin{figure}[htb]
\centering
{\bf(a)}
\includegraphics[width=0.35\textwidth]{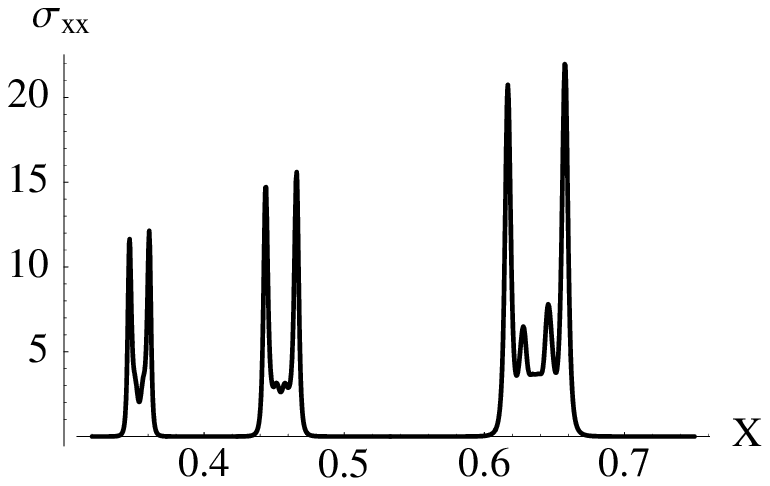}\\
\centering
{\bf(b)}
\includegraphics[width=0.35\textwidth]{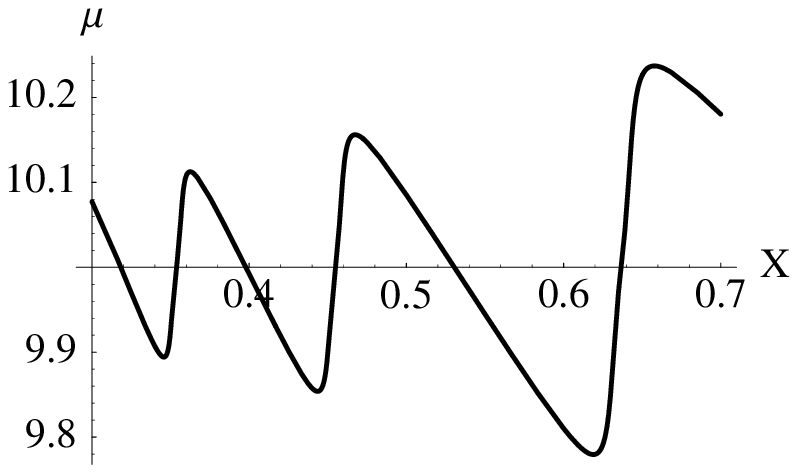}\\
\centering
{\bf(c)}
\includegraphics[width=0.35\textwidth]{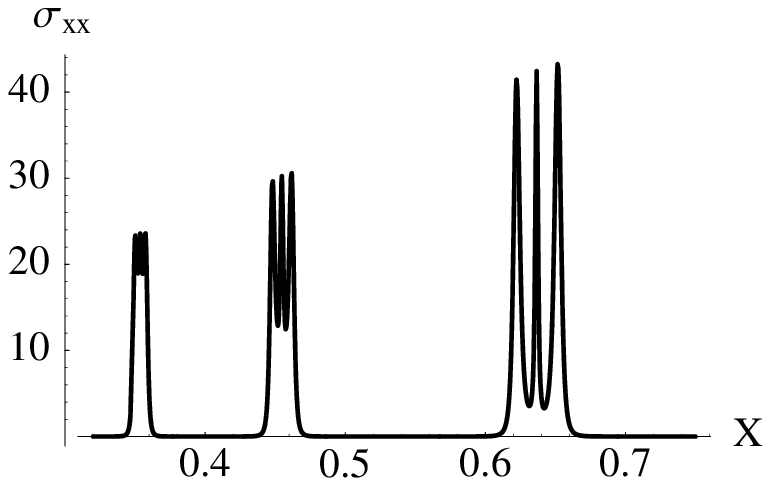}\\
\caption{The conductivity $\sigma_{xx}(X)$ given by Eq.(\ref{14})
(Fig.4a) in which the chemical potential $\mu(X)$ oscillates as
in Fig.4b (a direct sawtooth, see Eq.(\ref{17}) and text).
The choice of the units is the same as in Fig.2. The spin-splitting
parameter $s=0.093$, $E_{F}=10$, $\nu=0.06$, and  $X=\hbar\Omega$.
In Fig.4c $s=0$.}
\end{figure}
The spin-splitting parameter $s=2\pi\mu_{e}B/\hbar\Omega$ can be rewritten
in terms of the $g$-factor and the effective mass to the electron mass ratio
$s=\pi g (m^*/m)$. In GaAs $g\approx 0.44$ and $(m^*/m)\approx 0.068$ 
which yield  $s\approx 0.093$. This value gives a pronounced splitting in
the peaks in Fig.4a, but it is much less noticeable in Fig.4b for 
$\mu(B)$. The shape of peaks in the absence of splitting ($s=0$) is shown in
Fig.4c. For the direct sawtooth shape of the $\mu(B)$ the
peak-splitting is shown in Fig.5 in a more detail.
As explained in \cite{2} the difference in the shape of the 
$\sigma_{xx}(B)$ is because the equation for the split-peaks
positions has different number of the real and imaginary roots for the
direct and inverse shapes of the sawtooth function $\mu(B)$.
\begin{figure}[htb]
\centering
{\bf(a)}
\includegraphics[width=0.35\textwidth]{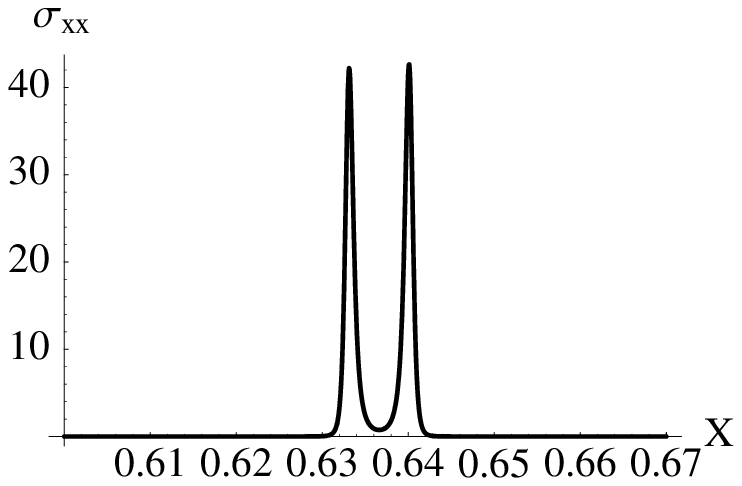}\\
\centering
{\bf(b)}
\includegraphics[width=0.35\textwidth]{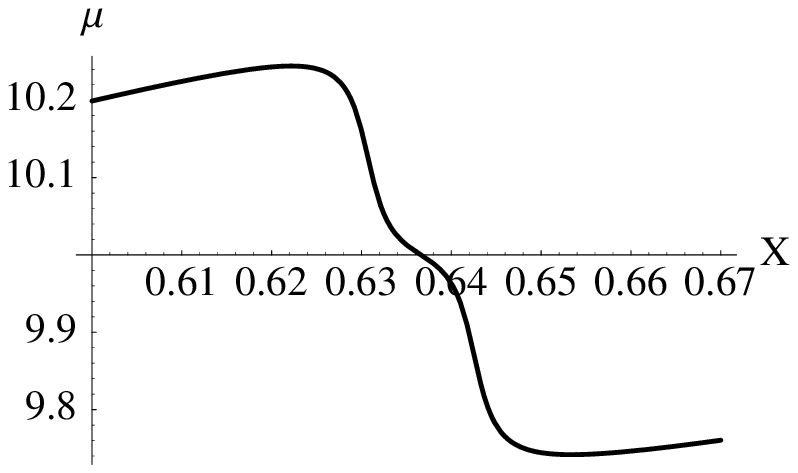}\\
\caption{$\sigma_{xx}(X)$ and $\mu(X)$ - direct sawtooth,
$\nu=0.03$.}
\end{figure}

In conclusion, the considered model of the hopping conductivity describes
the different regimes in the diagonal conductivity $\sigma_{xx}$, as
stated above in (i)-(iv). It also explains why the VRH exponent in the
conductivity $\sigma_{xx}$ corresponds to a 1D system while the
system in question is a 2D.  It is worth to note that
the peak-split shape in Fig.4(a) is typical for the IQHE conductors
with the high mobility of electrons. The approach developed is open for
the usage of specific models for the localization (see
Eq.(\ref{10})). The role of the quantum term in the $\sigma_{xx}$ is
similar to that considered in \cite{2,12,13} for the conductivity
across the layers in organic conductors.

The work was supported by INTAS, project
INTAS-01-0791. The author is deeply gratitude to P. Fulde and 
S. Flach for the hospitality at MPIPKS in Dresden.

\newpage

\end{document}